\documentclass[superscriptaddress,pre,showpacs,twocolumn,aps]{revtex4-1}
\usepackage{graphicx,color}
\usepackage{amssymb,amsmath,amsthm}
\usepackage{MnSymbol}

\newcommand{\reff}[1]{(\ref{#1})}
\DeclareMathOperator{\Tr}{Tr}

\begin{document}
\title{Casimir effect in swimmer suspensions}
\author{C. Parra-Rojas} 
\affiliation{Departamento de F\'isica, Facultad de Ciencias F\'isicas y 
Matem\'aticas, Universidad de Chile, Casilla 487-3, Santiago, Chile}
\affiliation{Theoretical Physics Division, School of Physics and Astronomy, The University of Manchester, Manchester M13 9PL, UK}

\author{R. Soto}
\affiliation{Departamento de F\'isica, Facultad de Ciencias F\'isicas y 
Matem\'aticas, Universidad de Chile, Casilla 487-3, Santiago, Chile}
\date{\today}

\begin{abstract}

We show that the Casimir effect can emerge in microswimmer suspensions. In principle, two effects conspire against the development of Casimir effects in swimmer suspensions. First, at low Reynolds number, the force on any closed volume vanishes, but here the relevant effect is the drag by the flow produced by the swimmers, which can be finite. Second, the fluid velocity and the pressure are linear on the swimmer force dipoles, and averaging over the swimmer orientations would lead to a vanishing effect. However, being the suspension a discrete system, the noise terms of the coarse grained equations depend on the density, which itself fluctuates, resulting in effective non-linear dynamics. Applying the tools  developed for other non-equilibrium systems to general coarse grained equations for swimmer suspensions, the Casimir drag is computed on immersed objects, and it is found to depend on the correlation function between the rescaled density and dipolar density fields.
By introducing a model correlation function with  medium range order, explicit expressions are obtained for the Casimir drag on a body. When the correlation length is much larger than the microscopic cutoff, the average drag is independent of the correlation length, with a range that depends only on the size of the immersed bodies.

\end{abstract}
\maketitle

\section{Introduction}

Microscopic swimmer suspensions constitute an interesting playground for non-equilibrium physics. Energy is continuously taken from the nutrients dissolved in the solution and used to produce directed motion. As an effect of the mutual interaction, swimmers present coherent motion with features similar to turbulence when the suspension is considered as an effective fluid~\cite{Cisneros,Wensink}. By their motion, swimmers also agitate the fluid  and it has been observed that this fluid agitation induces enhanced diffusion~\cite{Wu2000,Valeriani,Mino2010,Lin2011,Zaid2011} and generates directed motion~\cite{Sokolov,DiLeonardo}.  From a mechanical point of view, swimmers are autonomous objects and, therefore, the total force acting on them vanishes. Consequently, the net force exerted on the fluid vanishes as well and, at first order, swimmers can be modeled as force dipole tensors. Depending on whether the dipole is tensile or contractile, the swimmers are classified as pushers or pullers, respectively~\cite{HernandezOrtiz2005, Saintillan2007, Baskaran2009}. In the first category we find bacteria like  \emph{Escherichia coli}, while algae like  \emph{Chlamydomonas reinhardtii}  belong to the second category. 

Swimmer suspensions present high fluctuations in particle density and also in the orientation field when they align in domains. It has been argued that giant density fluctuations develop as a consequence of the coupling with the orientation field in presence of self-propulsion \cite{Giant1,Giant2,Giant3}. Also, the orientation field shows long wavelength fluctuations in the form of Goldstone modes that are extremely soft \cite{Goldstone1,Goldstone2}. 
Thanks to the fluctuations in orientation, swimmer suspensions---even in the ordered phase---do not show long-range order. At large scales they look homogeneous and isotropic. 
It is interesting to question whether these large fluctuations can generate macroscopic phenomena. When the fluctuating fields are limited to some modes due to the presence of boundary conditions---for example, due to immersed bodies---the Casimir effect can appear. 
The presence of this effect can have important effects on the motion and self-assembly of immersed objects. 

Normally, when two large bodies are immersed in the fluctuating medium, there is a pressure difference between the region bounded by the bodies and the exterior region, giving rise to a force. This pressure difference emerges as a result of the renormalization of the pressure by the fluctuations \cite{Soto2007}. 
Microswimmers are governed by hydrodynamics at low Reynolds numbers where  the generated stresses are linear in the force dipole intensity. As as result, when averaging over the different swimmer orientations it is expected that no renormalization of the pressure or fluid flow is possible, leading at first sight to a vanishing Casimir effect. In this article we will investigate the emergence of Casimir effect and show that, thanks to the large density fluctuations Casimir effects can develop. Indeed, the coarse grained  equations that describe the dynamics of the suspension have noise terms that are proportional to the square root of the density, implying that the stochastic equations are non-linear \cite{Dean,Chavanis}. A second, minor concern is that at low Reynolds number, the total force over any body immersed in the fluid adds up to zero \cite{Kim,Happel}. However, in this Stokes regime it is not the force but the drag on the immersed bodies that is the relevant quantity that will turn out to be finite due to the fluctuations. Recently, it has been proposed that a Casimir-like effect can be originated in the momentum transfer at swimmer-walls collisions (steric interactions) \cite{Reichhardt}, which can be a complementary mechanism of the one proposed here. 

The article is organized as follows. Section \ref{sec.flucthydro} presents the fluctuating description of the swimmer suspension and how the average drag is obtained in terms of coarse grained fields. In Section \ref{sec.casimir} it is shown that the noise terms, which are non-linear, imply that the primary fluctuating fields are non-Gaussian. By making a change of variables we generate a framework that allows us to compute the Casimir drag in terms of correlation functions. Section \ref{sec.mech2} presents a complementary mechanism that also generates a Casimir effect that is due to  non-linear couplings that emerge near the ordering transition. The drag generated by this mechanism is obtained by performing similar computations to the principal case under study. To analyze both cases, a model for the relevant correlations that are needed for the calculation is introduced in Section \ref{sec.mediumrange}, giving rise to explicit expressions for the drag. Finally, conclusions and perspectives are presented in Section \ref{sec.conclusions}.

\section{Fluctuating description of an active suspension} \label{sec.flucthydro}
We consider a suspension of swimmers in a fluid in three dimensions, that we assume to be homogeneous and isotropic on the large scale. Each swimmer is described by its director $\hat{\mathbf{n}}$, which points along its direction of motion. Axisymmetric swimmers are characterized by a force dipole tensor acting on the fluid
\begin{align}
S_{jk}=\sigma_0 n_jn_k,
\end{align}
where $\sigma_0$ is the dipole intensity, negative for pushers and positive for pullers. 
The effect of the force dipole is to move the fluid around the swimmer, with a velocity field that is obtained by solving the Stokes equations, valid at low Reynolds number. Also, each swimmer generates a stress field (pressure plus viscous stresses), but as indicated on the introduction the total force on any close object vanishes and therefore the stresses cannot produce a Casimir effect. We concentrate then on the velocity field that, for a swimmer located at $\mathbf{r}_0$, is given by
\begin{align}
u_i(\mathbf{r})=J_{ij,k}(\mathbf{r}-\mathbf{r}_0)S_{jk} ,\label{eqn:Umicro}
\end{align}
where $J_{ij,k}$ is the gradient of the Oseen tensor along the direction $k$
\begin{align}
J_{ij,k}(\mathbf{r}) &= \frac{1}{8\pi\eta r^3}\left(\delta_{ik}r_j + \delta_{jk}r_i - \delta_{ij}r_k - 3\frac{r_ir_jr_k}{r^2} \right) 
\end{align}
and summation over repeated indices is assumed~\cite{Kim,Happel}. 

When several swimmers are placed in the fluid, by the linearity of the Stokes equations, the resulting flow field is the sum of the effects produced by each swimmer. In a suspension  of $N$ swimmers  placed in a volume $V$, the dipolar density is defined as
\begin{align}
s_{jk}(\mathbf{r})&=\sum\limits_{\alpha=1}^N\delta\left(\mathbf{r}-\mathbf{r}^{\alpha}\right)S_{jk}^{\alpha},\label{eqn:Stress}
\end{align}
where $S_{jk}^{\alpha}=\sigma_0 n_j^{\alpha}n_k^{\alpha}$ is the dipolar tensor of the $\alpha$-th swimmer located at $\mathbf{r}^{\alpha}$ and we have assumed that all swimmers have the same dipolar intensity $\sigma_0$. Note that, since $\hat{\mathbf{n}}$ is a unit vector $\Tr s_{jk}(\mathbf{r}) = \sigma_0 \rho(\mathbf{r})$, where $\rho$ is the local number density of swimmers. In terms of the dipolar density, the velocity field is 
\begin{align}
u_i(\mathbf{r})&=\int \limits_V \! d^3 r'\, J_{ij,k}(\mathbf{r}-\mathbf{r}')s_{jk}(\mathbf{r}'). \label{eqn:Vel}
\end{align}

We are interested in the ensemble average $\langle u_i(\mathbf{r}) \rangle$, which  depends on $\left\langle s_{jk} \right\rangle$. To give its statistical properties, we consider a coarse grained description for the dominant fields which are the swimmer density $\rho$, the polar density field ${\bf p}$, which is related to the average director $\boldsymbol{\tau}={\bf p}/\rho$, and the already described dipolar density tensor field $\boldsymbol{s}$. The coarse grained descriptions adopt the form of fluctuating hydrodynamic equations that in general are coupled equations of the form~\cite{Dean,Chavanis}
\begin{align}
\partial_t \rho &= g_1[\rho, \boldsymbol{p},  \boldsymbol{s}] +\sqrt{\rho} \eta, \label{eqn:GeneralEQ1}\\
\partial_t p_{i} &= g_2[\rho, \boldsymbol{p},  \boldsymbol{s}] + \sqrt{\rho}\xi_{i}, \label{eqn:GeneralEQ2}\\
\partial_t s_{jk} &= g_3[\rho, \boldsymbol{p},  \boldsymbol{s}] + \sqrt{\rho}\zeta_{jk}, \label{eqn:GeneralEQ3}
\end{align}
where $g_n$ are functionals of the fields, which depend on the symmetries, conservation
laws, interactions and models for activity ~\cite{Goldstone1,Igor07,Marchetti08,Bertin09,Marchetti12,Igor12}. The noise terms are modeled---as it is usually done---to be white which, under isotropic conditions, are characterized by the following statistical properties
\begin{align*}
&\left\langle \eta(\mathbf{r},t)\right\rangle = 
\left\langle \xi_i(\mathbf{r},t)\right\rangle =  
\left\langle \zeta_{ij}(\mathbf{r},t)\right\rangle = 0\\
&\left\langle\eta (\mathbf{r},t) \xi_i(\mathbf{r}',t')\right\rangle = 
\left\langle\eta(\mathbf{r},t) \xi_{i}(\mathbf{r}',t')\right\rangle = 
\left\langle\xi_i(\mathbf{r},t) \zeta_{jk}(\mathbf{r}',t')\right\rangle =0\\
&\left\langle \eta(\mathbf{r},t)\eta(\mathbf{r}',t')\right\rangle = \Gamma_1 \delta(\mathbf{r}-\mathbf{r}')\delta(t-t')\\
&\left\langle \xi_{i}(\mathbf{r},t)\xi_{j}(\mathbf{r}',t') \right\rangle = \Gamma_2 \delta_{ij} \delta(\mathbf{r}-\mathbf{r}')\delta(t-t')\\
&\left\langle \zeta_{ij}(\mathbf{r},t)\zeta_{kl}(\mathbf{r}',t') \right\rangle = \left[\Gamma_3 \delta_{ij}\delta_{kl} 
+\Gamma_4 (\delta_{ik}\delta_{jl} 
+ \delta_{il}\delta_{jk}) \right]\\
& \phantom{\left\langle \zeta_{ij}(\mathbf{r},t)\zeta_{kl}(\mathbf{r}',t') \right\rangle = }
 \times \delta(\mathbf{r}-\mathbf{r}')\delta(t-t').
\end{align*}
We remark that the noise intensities depend on density because the system is particulate. Indeed, the fluctuations are originated in the displacements and interactions of the swimmers, events that in the limit of large numbers are described by Poissonian statistics leading to deviations that are proportional to the square root of the number of individuals \cite{Dean,Chavanis}. 

\section{Casimir effect} \label{sec.casimir}
Models with explicit expressions for Eqs. (\ref{eqn:GeneralEQ1}-\ref{eqn:GeneralEQ3}) have been described in several cases~\cite{Igor07,Marchetti08,Bertin09,Marchetti12,Igor12}. Here, without going into specific details of these models, we will show that they generally present Casimir effects. 
As the equations for the fields are coupled and the noise terms enter multiplicatively, in general  the fluctuations of the dipole field $s_{jk}$ around its equilibrium value will not be linear in the noise. Therefore, its stationary probability distribution function will not be Gaussian and its average will not vanish, giving rise to Casimir effects. 
Note that in particulate systems the noise term are always multiplicative. However, in many cases the systems are approximately incompressible and this non-linearity is irrelevant. Swimmer suspensions, on the contrary, present large fluctuations and this dependence cannot be neglected. 

The coupling with the noise terms is made linear---additive noise---by means of defining new fields $\phi$, $ \boldsymbol{\psi}$, and $\boldsymbol{\chi}$ such that $\rho= \rho_0 + \phi$, $p_i=\sqrt{\rho}\psi_i$, and  $s_{jk}=\sqrt{\rho}\left[ \frac{\sigma_0}{3}\sqrt{\rho_0}\delta_{jk} + \chi_{jk}\right]$. Replacing these expressions in  Eqs. (\ref{eqn:GeneralEQ1}-\ref{eqn:GeneralEQ3})  and linearizing the equations in the new fields we obtain
\begin{align}
\partial_t \phi &= \tilde{g}_1[\phi,  \boldsymbol{\psi}, \boldsymbol{\chi}]+\eta ,\label{eq.linear1}\\
\partial_t \psi_i &= \tilde{g}_2[\phi,  \boldsymbol{\psi}, \boldsymbol{\chi}]+\xi_i ,\label{eq.linear2}\\
\partial_t \chi_{jk} &= \tilde{g}_3[\phi,  \boldsymbol{\psi}, \boldsymbol{\chi}] + \zeta_{jk} ,\label{eq.linear3}
\end{align}
where now the noise terms enter additively and the functionals $\tilde{g}_n$ are linear. Consequently, now all the fluctuating fields have Gaussian statistics with zero mean.

In terms of the new fields, the average dipolar density is
\begin{align}
\langle s_{jk}(\mathbf{r}) \rangle = \frac{\rho_0\sigma_0}{3} \delta_{jk} + \frac{1}{2\sqrt{\rho_0}} \langle \phi(\mathbf{r}) \chi_{jk}(\mathbf{r}) \rangle,
\end{align}
which is now quadratic on the linearly fluctuating fields; hence, its average will be generally different from zero.
The isotropic part of the stress does not contribute to the velocity field, property that is represented in Stokesian flows by the relation $J_{ij,k}\delta_{jk}=0$. Therefore, we are left to compute the cross correlation $\langle \phi \chi_{jk} \rangle$. 
The Casimir effect emerges in non-equilibrium systems because the value of this cross correlation depends on the geometry. In particular it is modified by the presence of immersed bodies that introduce boundary conditions on the fluctuating fields.  Here, the potential Casimir effect would consist on the drag of the immersed bodies and therefore they should not be considered as fixed objects. However, if the drag velocity is small, we can consider that the swimmers see the intruders as  impenetrable bodies. Consequently they impose a non-flux boundary condition for the swimmer density that translates into a non-flux boundary condition for $\phi$. We do not have a natural boundary condition for the  dipolar density $\boldsymbol{s}$ and the associated field $\boldsymbol{\chi}$, which should be obtained from kinetic models that include  swimmer-object interactions (for example \cite{Marconi}). For lack of these models we consider for simplicity that there are also non-flux boundary conditions for $\boldsymbol{\chi}$, but other boundary conditions can be studied in an analogous way, leading to similar results, although the sign of the effect may be reversed as it happens in the critical Casimir effect \cite{CriticalCasimir}.

\begin{figure}[htb]
\includegraphics[width=.8\columnwidth]{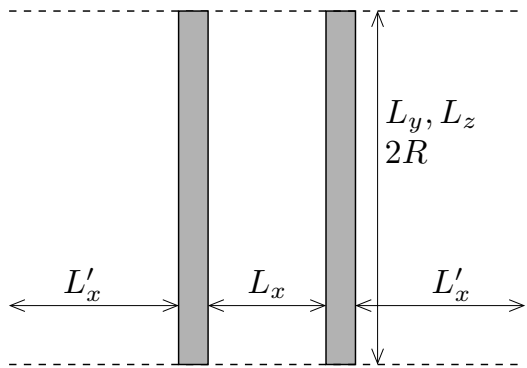}
\caption{Geometry used for the calculation of the Casimir drag. The bodies confine an active suspension in a region of size $L_x  \times L_y \times L_z$. There are non-flux boundary conditions at $x=0$ and $x=L_x$, while the fields are periodic in $y$ and $z$. In Eq. \reff{eqn:MeanV} the cross sections of the bodies are modeled as circles of radius $R$. The total drag on the bodies results from the subtraction of the drag produced by the region inside (separation $L_x$) and outside (separation $L_x'$) the bodies. To obtain the final expression \reff{Eq:finaldrag} the limit $L_x'\to\infty$ is considered.}
\label{fig.geom}
\end{figure}

To perform the calculation we consider a geometry and a protocol similar to the one used in Ref. \cite{Cattuto} (Fig. \ref{fig.geom}). That is, two equal bodies are immersed in the fluid. If the separation between the bodies is small compared to their size, the volume in between can be modeled to have non-flux boundary conditions on the bodies' surfaces and periodic boundary conditions in the other direction. The activity in the region inside will generate a drag on the objects that should be subtracted to a similar drag on the other side of the objects. To make an illustrative calculation of the Casimir drag and, specially, to show that it gives non-vanishing results we will simplify the geometry to that of a parallelepiped.
We proceed in a similar way as in Refs. \cite{Soto2007, SotoGranular}, considering a volume $V=L_x \times L_y \times L_z$ with non-flux boundary conditions for $\phi$ and $\boldsymbol{\chi}$ at $x=0$ and $x=L_x$, while the fields are periodic in $y$ and $z$. Using these boundary conditions the fluctuating density field is expanded as
\begin{align}
\phi(\mathbf{r},t) &= V^{-1}\sum\limits_{k_x} \sum\limits_{k_y}\sum\limits_{k_z}\phi(\mathbf{k},t) \cos (k_xx)e^{ik_yy}e^{ik_zz},
\end{align}
where $k_x=\pi n_x/L_x$, $k_y=2\pi n_y/L_y$, $k_z=2\pi n_z/L_z$, $n_x=0,1,2,\ldots$; $n_y,n_z=\ldots,-1,0,1,\ldots$. Analogous expressions are used for $\chi_{ij}$. 

The density-dipole correlation in Fourier space results in
\begin{align}
\left\langle \phi (\mathbf{k})\chi_{jk}(\mathbf{q})\right\rangle &= \gamma_{k_x} V {G}_{jk}(\mathbf{k}) \hat{\delta}_{\mathbf{k},\mathbf{q}}, 
\end{align}
where ${G}_{jk}(\mathbf{k})$ is the density-dipole structure factor in the bulk that can be obtained by solution of the coarse grained equations \reff{eq.linear1}-\reff{eq.linear3} with full periodic boundary conditions. The prefactor $\gamma_{k_x}$ ($\gamma_{k_x} =1/2$ if $k_x=0$ and $\gamma_{k_x}=1$ if $k_x\neq 0$) and the modified Kroenenker delta  
$\hat{\delta}_{\mathbf{k},\mathbf{q}} = \delta_{k_x,q_x}\delta_{k_y,-q_y}\delta_{k_z,-q_z}$ appear from the use of the non-flux boundary conditions~\cite{Soto2007}.

Going back to real space, we compute 
\begin{align}
C_{jk}(\mathbf{r}) &= \left\langle \phi(\mathbf{r}) \chi_{jk}(\mathbf{r})\right\rangle \nonumber\\
&= V^{-1} \sum\limits_{\mathbf{k}} \! ^\prime  {G}_{jk}(\mathbf{k})\cos (k_xx)^2, 
\end{align}
where the prime in the sum indicates that the term $k_x=0$ is multiplied by 1/2 and it only depends on $x$ due to the periodic boundary conditions in $y$ and $z$. It is possible now to compute the velocity field 
\begin{align*}
\left\langle u_{i}(x)\right\rangle =& \frac{1}{2\rho_0^{1/2}} \int d^3r' J_{ij,k}(x-x',y',z') C_{jk}(x')\\
=& \frac{1}{2\rho_0^{1/2}V}  \sum\limits_{\mathbf{k}} \! ^\prime \int d^3r'  \cos^2 (k_xx') \nonumber \\
&   \times J_{ij,k}(x-x',y',z'){G}_{jk}(\mathbf{k}).
\end{align*}

For an isotropic system, the bulk structure factor can be generally expressed as
\begin{align}
{G}_{jk}(\mathbf{k}) &= A(k)\delta_{jk} + B(k)\frac{k_jk_k}{k^2} \label{tensorAB}
\end{align}
in terms of two scalar functions of the wavenumber.  Again the isotropic part does not contribute to the velocity field and we have
\begin{align}
J_{ij,k}(\mathbf{r}){G}_{jk}(\mathbf{k}) &= \frac{1}{8\pi\eta} B(k)\frac{r_i}{r^3}\left[1 - 3\frac{(\mathbf{r}\cdot \mathbf{k})^2}{r^2k^2} \right].
\end{align}

The swimmer suspension produces a net velocity field as an effect of the confinement and the non-linear fluctuations of the dipolar field. This velocity field generates a Casimir effect by noting that the velocity at the intruders surfaces does not vanish, i.e. they are dragged by the fluid. 
The relevant drag takes place in the $x$ direction, and has the form
\begin{align}
\left\langle u_{x}(x)\right\rangle =& \frac{1}{16 \pi\eta \rho_0^{1/2}V} \sum\limits_{\mathbf{k}} \! ^\prime \int d^3r' \cos^2(k_x x') B(k) \nonumber \\
&  \times \frac{l_x}{l^3} \left[1 - 3 \frac{\left({\bf l}\cdot{\bf k}\right)^2}{ l^2k^2}\right], \label{uxB}
\end{align}
where ${\bf l}=(x-x',y',z')$ and we recall that the integration in ${\bf r}'$ is over the position of the dipole sources while $x$ is the position where the field is evaluated. To compute the drag on the wall, the velocity field must be evaluated at its location ($x=0$) to be further averaged over the wall surface. Here we note a peculiarity of the Stokes flows: as a consequence of the incompressibility condition, when a point source is placed in a fluid,  the integrated velocity across an infinity surface vanishes identically \cite{Happel,Kim}. By linearity, a distribution of dipolar sources produces the same result. Then, if the velocity field \reff{uxB} were averaged over an infinite surface it would also vanish. However, the immersed bodies are finite. In order to achieve simple results, we consider an immersed body of circular cross section of radius $R$. The average velocity reads
\begin{align}
\left\llangle u_{x}(0)\right\rrangle &= -\frac{1}{16 \pi\eta \rho_0^{1/2}L_x} \sum\limits_{\mathbf{k}} \! ^\prime \int d x' \cos^2(k_x x') \nonumber \\
& \qquad \times B(k) \frac{x' \left(k^2-3 k_x^2\right)}{k^2\left(R^2+x'^2\right)^{3/2}},\label{eqn:MeanV}
\end{align}
where $\left\llangle \ldots\right\rrangle$ means an ensemble average over the noise and an average over the body surface.

\section{Complementary mechanism} \label{sec.mech2}
There is a second mechanism that can also generate a non-vanishing average of $\boldsymbol{s}$ and therefore induce a Casimir effect. 
Pusher suspensions (e.g. bacterial baths) and rod-like swimmers with steric interactions can present a polar transition where the average director field $\boldsymbol{\tau}$ is finite through a spontaneous symmetry breaking \cite{Goldstone1,Goldstone2}. Close to the transition, but still in the isotropic phase the density  and director fields become slow variables and the other fields are enslaved to them. Specifically, the dipole tensor field is found to be $s_{jk} = \sigma_0 \left[(1-\lambda \tau_l \tau_l)\rho\delta_{jk}/3 + \lambda\rho\tau_j\tau_k \right]$ with a positive dimensionless constant $\lambda$ \cite{aranson2006theory}. Therefore, apart from the isotropic term, it is quadratic in the fluctuating field.

Deep in the polar phase the average director field $\boldsymbol{\tau}$ has finite norm. But the system is not in a global symmetry broken state because the orientation has only a finite correlation length due to Goldstone mode fluctuations. Therefore, at large length scales the system is globally isotropic and the analysis we have performed can be applied here. In the perfectly locally ordered case ($|\boldsymbol{\tau}|\approx 1$), the dipole tensor is $s_{jk} = \rho_0\sigma_0 \tau_j \tau_k$, being also quadratic in the fluctuating field. 

In an isotropic medium the average $\langle \tau_i\tau_j\rangle$ can be written also like in \reff{tensorAB}. The derivation then follows in exactly the same way as in the previous section with a induced Casimir drag that is given by the expression \reff{eqn:MeanV}, albeit with a different prefactor. 
Being the expressions identical we continue the discussion using the notation of the mechanism described in the previous section.

\section{Model with medium range order} \label{sec.mediumrange}

To proceed with the calculation of the Casimir effect, we need to provide a model for the density-dipole structure factor $B(k)$. To our knowledge, this function has not been measured in swimmer suspensions. Here we consider a model with medium range order that can describe situations were the suspension is showing structures with a finite correlation length.

The tensorial structure factor ${G}_{ij}(\mathbf{k})$ needs to meet two conditions: it must vanish for $k\to \infty$ and it must be single-valued at the origin. From this second requirement, it follows that $B(0)=0$, and we can write $B(k)=k^2 \tilde{B}(k)$. The simplest assumption we can make is that $\tilde{B}$ is characterized by a single correlation length $k_0^{-1}$ that, eventually, can diverge at a critical point as it could happen in a swarming phase or in other phases with collective order~\cite{Goldstone1}. Following our approach in \cite{Parra2013} we take, therefore, ${B}$ as a simple rational function with a correlation length $k_0^{-1}$
\begin{align}
B(k) &= \sigma_0 \Gamma \frac{k^2}{\left({k_0}^2+k^2\right)^2}. \label{Bmedium}
\end{align}  
The prefactor $\Gamma$ is a measure of the correlation intensity, which will be a function of the noise intensities $\Gamma_{1,2,3,4}$, and we have factored out the dependence on the dipole strength. The sign of $\Gamma$ depends on the particular model that describes the swimmer suspension. A stochastic extension of the kinetic model presented in Ref. \cite{aranson2006theory} predicts a positive value \cite{ParraToBePublished}. In the case of the second mechanism where the dipolar density is enslaved to the director field, the sign of $\Gamma$ is given by the sign of $[3\langle(\hat{\bf k}\cdot\boldsymbol{\tau}({\bf k}))^2\rangle-\langle |\boldsymbol{\tau}({\bf k})|^2\rangle]$, which is positive if the director field develops longitudinal structures and negative if vortex-like structures are formed.

More accurate models, obtained from experiments, discrete element simulations~\cite{Wensink} or continuos models~\cite{rheology} will change only qualitatively the picture below if they are characterized by a single correlation length. More complex  models, with different scaling at large distances, should be worked out separately.

Once Eq.~\reff{Bmedium} is substituted into \reff{eqn:MeanV}, the sums over the transverse wavevectors $k_y$ and $k_z$ can be replaced by integrals when $R$ is large compared with $L_x$. To integrate, we introduce a cutoff for large wavevectors $2\pi/a$ in order to take into account that the continuous model is valid up to the microscopic length $a$, resulting in the expression
\begin{widetext}
\begin{align}
\left\llangle u_{x}(0)\right\rrangle &= - \frac{\tilde{R}^2\sigma_0\Gamma}{32\pi^2\eta \tilde{L}_x\rho_0^{1/2}}  \sum\limits_{\tilde{k}_x} \! ^\prime \int d \tilde{x}'  \frac{\cos^2 (\tilde{k}_x\tilde{x}')\tilde{x}'}{\left(\tilde{R}^2+\tilde{x}'^2\right)^{3/2}}\left[\tanh ^{-1}\left(\frac{2\pi^2}{2\pi^2+\tilde{a}^2\left(1+\tilde{k}_x^2\right)}\right)-\frac{2\pi^2\left(1+3\tilde{k}_x^2\right)}{\left(1+\tilde{k}_x^2\right)\left(4\pi^2+\tilde{a}^2\left(1+\tilde{k}_x^2\right)\right)}\right] ,\label{Eq:uxmodel}
\end{align}
\end{widetext}
where $\tilde{L}_x=k_0L_x$, $\tilde{R}=k_0R$, $\tilde{x}'=k_0x'$, $\tilde{k}_x=k_x/k_0$ and $\tilde{a}=k_0 a$. 

The most relevant case for the Casimir effect is when there is a finite correlation length, much larger than the microscopic cutoff, in  which case $\tilde{a}\ll 1$. If in Eq. \reff{Eq:uxmodel} we  use $\cos^2(\tilde{k}_x \tilde{x}') = 1/2 + \cos(2\tilde{k}_x \tilde{x}')/2$ the constant contribution goes as $1/\tilde{a}$ for small $\tilde{a}$, while the oscillatory one goes as $\log \tilde{a}$. Therefore, in the relevant regime, $\tilde{a}\ll 1$, we can consider only the constant contribution that is the dominant one. The sums can be done numerically and the results, computed for the cases $\tilde{R}\ll 1$, $\tilde{R}\sim 1$ and $\tilde{R}\gg 1$, are all well fitted by the expression
\begin{align}
\left\llangle u_{x}(0)\right\rrangle &= - \frac{\tilde{R}^2\sigma_0\Gamma}{32\pi^2\eta \rho_0^{1/2}} \frac{c_0 \tilde{L}_x^2}{\tilde{a}\tilde{R}(\tilde{L}_x^2 +c_1 \tilde{R}^2)},
\end{align}
where $c_0=0.29$ and $c_1=1.62$. For large distances compared to the intruders' size ($\tilde{L}_x\gg \tilde{R}$) the expression saturates to a constant value, while for small distances it grows like $\tilde{L}_x^2$.

As usual when considering Casimir effects, the total drag on a surface is obtained by the subtraction of the drag generated in the region at one side of the intruder with the drag generated on the other side (see Fig. \ref{fig.geom}). A simple case corresponds to considering that the region to the left of the body is large ($\tilde{L}_x\gg \tilde{R}$) such that the asymptotic expression can be used, resulting in
\begin{align}
u^{\rm total}_{x} = \frac{\sigma_0\Gamma}{32\pi^2\eta \rho_0^{1/2}} \frac{c_0  R}{ \tilde{a}}
\left[1-  \frac{L_x^2}{L_x^2 +c_1 R^2}  \right] \label{Eq:finaldrag}
\end{align}
that, remarkably, does not depend on the correlation length.
The range of the Casimir drag scales with the size of the immersed body. This property is an effect of the long range effect of the hydrodynamic interactions. A similar result was obtained for the fluid velocity-velocity correlation function in a swimmer suspension, even if the swimmer correlations were short range \cite{Parra2013}. If $\Gamma\sigma_0>0$ the Casimir drag is positive, meaning that immersed objects are attracted, while in the opposite case the intruders are repelled. The  stochastic extension of the kinetic model for swimmers predicts $\Gamma>0$ \cite{aranson2006theory,ParraToBePublished}, therefore a suspension of 
pushers ($\sigma_0<0$) would lead to a repulsive drag.
Note that the drag depends on the cutoff length $a$. Normally, in the Casimir effect in quantum electrodynamics or critical fluids this is not the case. In non-equilibrium systems the results depend on the specific system under study, with cases that depend on the cutoff \cite{Cattuto} while others are cutoff-independent \cite{Soto2007}. Nevertheless, this is not a serious issue because here there is a natural cutoff given by the swimmer size and there is no \emph{a priori} reason to expect that there is an ultraviolet regularization.

\section{Conclusions} \label{sec.conclusions}

We have shown that a Casimir effect is present in low Reynolds number swimmer suspensions. It consists on an average drag over immersed objects which result from the fluctuating dipolar density field. Although the deterministic dynamics at low Reynolds number is linear, the stochastic dynamics that governs fluctuations is non-linear because the noise intensities are proportional to the square root of the density, which is also a fluctuating field. Changing variables to new fields where now the linear fluctuations are Gaussian, the drag on an immersed body turns out to be quadratic function of the new fields. The average drag is susceptible to have a contribution of the different modes of the fluctuating fields which result in a Casimir effect when the allowed modes are different on both sides of the immersed objects. 
The intensity of the Casimir drag depends on the correlation function of the rescaled density and dipolar density tensor fields. These correlations have not been measured in experiments or discrete element simulations and we propose a simple model with medium range order for a medium that is isotropic and homogeneous at the large scale. The resulting drag range depends on the body size and separation, but not on the correlation length, which is a result of the long range interactions in Stokes flows. 

In order to make more quantitative predictions, measurements of the relevant correlation functions are needed, which could be done for example by confocal microscopy methods as in Ref. \cite{confocal} where it was possible to track simultaneously the position and orientation of microscopic objects. Also, a proper modeling of the dipolar density boundary condition on immersed bodies is needed as other boundary conditions than the one used here (non-flux for the density and dipolar density tensor) could change the sign of the effect, as it has been observed for example in critical Casimir forces \cite{CriticalCasimir}.

In non-equilibrium systems the Casimir effect can lead to new phenomena, as compared to its equilibrium counterparts. Notably, there is the possibility that a single immersed object of asymmetric shape can experience a drag on its own leading to self-propulsion originated in fluctuation-induced phenomena \cite{Buenzli,QEDNonEq}. In principle, there is no reason \emph{a priori} to exclude this possibility, but precise calculations or experiments would need to be performed to confirm this.

\section*{Acknowledgment}
This research is supported by Fondecyt Grant No. 1140778. C.P.-R. acknowledges the support of a Becas Chile CONICYT No. 72140425.

\end{document}